\documentclass[12pt,preprint]{emulateapj}

\shorttitle{Spectropolarimetry of Optically Dull AGNs}
\shortauthors{Trump et al.}

\begin{document}

\title{Spectropolarimetric Evidence for Radiatively Inefficient
  Accretion in an Optically Dull Active Galaxy\altaffilmark{1}}

\author{
  Jonathan R. Trump,\altaffilmark{2}
  Tohru Nagao,\altaffilmark{3}
  Hiro Ikeda,\altaffilmark{3}
  Takashi Murayama,\altaffilmark{4}
  Christopher D. Impey,\altaffilmark{5}
  John T. Stocke,\altaffilmark{6}
  Francesca Civano,\altaffilmark{7}
  Martin Elvis,\altaffilmark{7}
  Knud Jahnke,\altaffilmark{8}
  Brandon C. Kelly,\altaffilmark{7}$^,$\altaffilmark{9}
  Anton M. Koekemoer,\altaffilmark{10}
  and Yoshi Taniguchi\altaffilmark{4}
}

\altaffiltext{1}{
  Based on observations with the Subaru Telescope, operated by the
  National Astronomical Observatory of Japan.
\label{cosmos}}

\altaffiltext{2}{
  University of California Observatories/Lick Observatory, University
  of California, Santa Cruz, CA 95064
\label{UCO/Lick}}

\altaffiltext{3}{
  Research Center for Space and Cosmic Evolution, Ehime University,
  2-5 Bunkyo-cho, Matsuyama 790-8577 Japan
\label{Ehime}}

\altaffiltext{4}{
  Astronomical Institute, Graduate School of Science, Tohoku
  University, Aramaki, Aoba, Sendai 980-8578 Japan
\label{Tohoku}}

\altaffiltext{5}{
  Steward Observatory, University of Arizona, 933 North Cherry Avenue,
  Tucson, AZ 85721
\label{Arizona}}

\altaffiltext{6}{
  Center for Astrophysics and Space Astronomy, Department of
  Astrophysical and Planetary Sciences, UCB-389, University of
  Colorado, Boulder, CO 80309
\label{Colorado}}

\altaffiltext{7}{
  Harvard-Smithsonian Center for Astrophysics, 60 Garden Street,
  Cambridge, MA 02138
\label{CfA}}

\altaffiltext{8}{
  Max Planck Institut f\"ur Astronomie, K\"onigstuhl 17, D-69117
  Heidelberg, Germany
\label{Max Planck 2}}

\altaffiltext{9}{
  Hubble Fellow
\label{Hubble}}

\altaffiltext{10}{
  Space Telescope Science Institute, 3700 San Martin Drive, Baltimore,
  MD 21218
\label{STScI}}

\def\etal{et al.}
\newcommand{\Ha}{\hbox{{\rm H}$\alpha$}}
\newcommand{\Hb}{\hbox{{\rm H}$\beta$}}
\newcommand{\OIII}{\hbox{[{\rm O}\kern 0.1em{\sc iii}]}}
\newcommand{\NII}{\hbox{[{\rm N}\kern 0.1em{\sc ii}]}}
\newcommand{\NaI}{\hbox{{\rm Na}\kern 0.1em{\sc i}~{\rm D}}}

\begin{abstract}

We present Subaru/FOCAS spectropolarimetry of two active galaxies in
the Cosmic Evolution Survey.  These objects were selected to be
optically dull, with the bright X-ray emission of an AGN but missing
optical emission lines in our previous spectroscopy.  Our new
observations show that one target has very weak emission lines
consistent with an optically dull AGN, while the other object has
strong emission lines typical of a host-diluted Type 2 Seyfert galaxy.
In neither source do we observe polarized emission lines, with
3$\sigma$ upper limits of $P_{BLR} \lesssim 2\%$.  This means that the
missing broad emission lines (and weaker narrow emission lines) are
not due to simple anisotropic obscuration, e.g., by the canonical AGN
torus.  The weak-lined optically dull AGN exhibits a blue polarized
continuum with $P = 0.78 \pm 0.07\%$ at $4400{\rm \AA} < \lambda_{\rm
rest} < 7200{\rm \AA}$ ($P = 1.37 \pm 0.16 \%$ at $4400{\rm \AA} <
\lambda_{\rm rest} < 5050{\rm \AA}$).  The wavelength dependence of
this polarized flux is similar to that of an unobscured AGN continuum
and represents the intrinsic AGN emission, either as synchrotron
emission or the outer part of an accretion disk reflected by a clumpy
dust scatterer.  Because this intrinsic AGN emission lacks emission
lines, this source is likely to have a radiatively inefficient
accretion flow.

\end{abstract}


\section{Introduction}

Deep X-ray surveys have revealed many bright X-ray point sources with
weak or no emission lines in their optical spectra
\citep[e.g.,][]{elv81,com02}.  Called X-ray bright optically normal
galaxies (XBONGs) or ``optically dull'' AGNs (the name adopted here),
these objects require an AGN to produce their high X-ray luminosities
but lack the broad or narrow emission line signatures of AGN
accretion.  Optically dull AGNs make up $\sim$15\% of luminous
($L_X>10^{42}$~erg/s) point sources in deep X-ray surveys, and
$\sim$25\% of those at $z<1$ \citep{tru09a,tro09,yan10}.  Three major
paradigms exist to explain their X-ray brightness and optical
dullness: (1) obscuration of both narrow and broad emission lines, (2)
dilution by host galaxy starlight, and (3) a physically distinct
accretion flow.  We explore the evidence for and against each paradigm
in turn.

The standard AGN unified model has had great success in using
obscuration to explain the differences between optically selected Type
1 (broad-line) and Type 2 (narrow-line) AGNs \citep{kro88,ant93}.  In
the simplest interpretation of this model, all AGNs have the broad
optical emission lines and strong UV/optical continua of Type 1 AGNs,
but along certain lines of sight these features are obscured within a
dusty ``torus'' a few parsecs from the black hole.  In an obscured
object the narrow emission lines remain visible because they are
excited beyond the obscuring material, and so the lack of a BLR and
weaker optical continuum of Type 2 AGNs are attributed to obscuration.
Similarly, \citet{com02} and \citet{civ07} suggested that optically
dull AGNs have the same physical engine as a Type 1 or Type 2 AGN, but
with additional obscuration within a few parsecs completely blocking
the ionizing continuum radiation from exciting even the narrow
emission lines.  \citet{rig06} instead suggest that optically dull
AGNs are obscured by extranuclear ($>$100 pc) gas and dust in the host
galaxy, blocking the narrow lines along the observer's line of sight.
No matter the physical location of the obscuring material, it must
preferentially absorb the optical emission, since optically dull AGNs
remain X-ray bright.  Indeed, more than half of optically dull AGNs
are relatively unobscured ($N_H<10^{22}$ cm$^{-2}$) in the X-rays
\citep{sev03,pag03}.  Preferentially obscuring the optical emission
while remaining X-ray unabsorbed would require extreme gas-to-dust
ratios not observed in other AGNs \citep{mai01}.

Optically dull AGNs could also be Type 2 AGNs with narrow emission
lines diluted by a bright host galaxy.  Indeed, \citet{mor02} showed
that many $z \sim 0$ AGNs would appear optically dull if observed at
$z \sim 1$, since the host galaxy would occupy more of the
spectroscopic slit or fiber and consequently overwhelm the AGN
emission lines.  HST/ACS images additionally show that many optically
dull AGN hosts at $z<1$ are edge-on \citep{rig06} or have a second
galaxy falling within the spectroscopic aperture \citep{tru09c}.
However 10-20\% of local (undiluted) AGNs are optically dull
\citep{laf02,hor05}.  And even after removing the host galaxy light by
decomposing the spectral energy distribution, $\sim$1/3 of optically
dull AGNs have anomalously high X-ray to optical flux ratios
\citep{tru09c}.

Another possibility is that optically dull AGNs have different
accretion physics due to low accretion rates \citep{yuan04}.  Models
have long predicted that an AGN with a low accretion rate ($\dot{m}
\equiv L_{\rm bol}/L_{\rm Edd} \lesssim 0.01$) will have a radiatively
inefficient accretion flow (RIAF) within some truncation radius $R_t$,
with $R_t$ defined as where the collisional cooling time is comparable
to the accretion time \citep{beg84,nar95}.  Beyond $R_t$, accretion
will remain in a standard geometrically thin and optically thick disk
with a thermal blackbody spectrum \citep[e.g.,][]{sha73}.  However
within $R_t$ there are too few collisions to couple the ions and
electrons and the gas becomes a two-temperature plasma.  The electrons
are cooled by bremsstrahlung, synchrotron, and Compton up-scattering,
while the ions remain at the virial temperature.  This means the flow
is geometrically thick and optically thin.  A RIAF then lacks much of
the optical/UV blackbody emission of an optically thick accretion
disk, and consequently cannot ionize and/or excite the broad or narrow
optical emission lines seen in other AGNs \citep{yuan04,tru11}.
Because the ions in the RIAF are only marginally bound, such AGNs
should also have strong outflows and be consequently radio luminous.
Optically dull AGNs, especially those with high X-ray to optical flux
ratios, are indeed observed to be have higher ratios of radio to
optical/UV luminosity than Type 1 AGNs \citep{tru09c,tru11}.  Much of
the optical and infrared light in RIAF AGNs may be synchrotron
radiation associated with the radio jet \citep{ho09}.

Each of these three paradigms has a different signature in
spectropolarimetry.  Anisotropic obscuration of the narrow emission
lines would leave telltale reflected emission lines in the polarized
spectrum \citep{nag04}.  On the other hand, if an optically dull AGN
is a Type 2 AGN diluted by a host galaxy, its polarized flux would be
equally diluted.  (A diluted optically dull AGN might exhibit
polarized broad emission lines like those seen in Type 2 AGNs, but the
host galaxy would probably overwhelm them to a non-detectable level.)
And if optically dull AGNs have RIAFs we might observe a featureless
polarized continuum, either from the synchrotron emission associated
with the stronger radio jet \citep[e.g.,][]{jan94,coh99} or because we
are viewing the naked, lineless continuum (from the disk beyond the
RIAF) reflected by a scattering surface in the host galaxy
\citep{ant85,ogl99,kis01}.  We summarize the polarization signatures
of each paradigm in Table \ref{tbl:paradigm}.

\begin{deluxetable}{ll}
\tablecolumns{7}
\tablecaption{Paradigms to Describe Optically Dull AGNs\label{tbl:paradigm}}
\tablehead{
  \colhead{Paradigm} & 
  \colhead{Spectropolarimetry Signature} }
\startdata
Obscuration & Polarized emission lines \\
Dilution & No detectable polarization (diluted by starlight) \\
RIAF & Polarized continuum but no polarized emission lines \\
\enddata
\end{deluxetable}

\section{Targets and Observations}

We used Subaru/FOCAS to observe the two brightest optically dull AGNs
of \citet{tru09c}, 095849+013220 and 100036+024929.  The objects have
multiwavelength observations from the Cosmic Evolution Survey
\citep[COSMOS,][]{sco07}, a survey based on a 1.7 deg$^2$ HST/ACS
treasury program \citep{koe07}.  Both targets have identifications and
redshifts from the XMM-COSMOS AGN spectroscopic campaign with
Magellan/IMACS \citep{tru09a}.  Despite their absorption line optical
spectra, they are confirmed as bona-fide AGN using their XMM data
\citep{cap09,bru10}: each has an X-ray luminosity of $L_{\rm
0.5-10~keV}>3 \times 10^{42}$ erg/s, requiring an AGN \citep{hor01}.
We show properties of 095849+013220 and 100036+024929 in Table
\ref{tbl:odagns}.

\begin{deluxetable*}{lcccccc}
\tablecolumns{7}
\tablecaption{AGN Target Properties\label{tbl:odagns}}
\tablehead{
  \colhead{RA+Dec} & 
  \colhead{z} & 
  \colhead{$i_{\rm AB}$} & 
  \colhead{$\log(L_X)$} & 
  \colhead{$f_{\rm AGN}$\tablenotemark{a}} & 
  \colhead{$X/O$\tablenotemark{b}} & 
  \colhead{$X/O$(AGN)\tablenotemark{c}} }
\startdata
095849.02+013219.8 & 0.361 & 18.94 & 44.83 & 0.45 &  1.1 &  1.4 \\
100036.21+024928.9 & 0.308\tablenotemark{d} & 18.77 & 43.03 & 0.45 & -1.0 & -0.6 \\
145658.70+221846.3\tablenotemark{e} & 0.258 & 19.59 & 42.18 & 0.20 & -0.9 & -0.2 \\
\enddata
\tablenotetext{a}{Fraction of $i$-band emission resulting from the
  AGN, as estimated from the template fitting of \citet{tru09c}.}
\tablenotetext{b}{$X/O = \log{f_X/f_O} = \log(f_{0.5-2~keV}) +
  i_{\rm AB}/2.5 + 5.352$}
\tablenotetext{c}{Ratio of X-ray to optical emission considering only
  the AGN component, including all of the X-ray emission and only
  $f_{\rm AGN}$ of the $i$-band emission.}
\tablenotetext{d}{Note that \citet{tru09a} and \citet{tru09c}
  incorrectly reported the redshift for 100036+024929 as $z=0.47$ with
  $z_{\rm conf}=3$ ($\sim$90\% confidence) because they used a
  Magellan/IMACS spectrum in which \Ha\ fell in a chip gap.  The
  Subaru/FOCAS spectrum presented here reveals that the correct
  redshift for this target is $z=0.308$.}
\tablenotetext{e}{This target is the cluster X-ray source MS1455 X-2,
  a BL Lac candidate discussed in the Appendix.}
\end{deluxetable*}

\citet{tru09c} used AGN plus galaxy fits to the optical photometry to
suggest that both of these targets are roughly 45\% AGN and 55\% host
galaxy in the $i$ band.  Table \ref{tbl:odagns} also gives the X-ray
to optical flux ratios, given by $X/O = \log{f_X/f_O} =
\log(f_{0.5-2~{\rm keV}}) + i_{\rm AB}/2.5 + 5.352$.  The second
target, 100036+024929, is a good candidate to be a normal AGN diluted
by a host galaxy because of its $X/O{\rm (AGN)}=-0.6$, fitting nicely
in the traditional ``X-ray AGN locus'' of $-1<X/O<1$ \citep{mac88}.
However the first target, 095849+013220, is anomalously X-ray bright
and optically dull, with $X/O=1.1$ and $X/O{\rm (AGN)}=1.4$.  Since
starlight dilution would increase the optical emission and decrease
the $X/O$ ratio, the high $X/O$ ratio for the first target means it
cannot be explained by starlight dilution.  Instead 095849+013220 is a
good candidate to host a RIAF \citep{tru09c}.

Spectropolarimetric observations of both targets were undertaken using
the Faint Object Camera and Spectrograph \citep[FOCAS,][]{kas02} on
the 8.2-m Subaru telescope.  Each target was observed for 6 hours of
total integration in a $0{\farcs}8$-wide center slit.  We used the R58
filter and the 150 l/mm grism, resulting in an observed spectral range
of 5800-10000\AA\ with $\sim$2.8 \AA/pixel resolution.  During
spectropolarimetry observations on FOCAS, a crystal quartz Wallaston
prism splits the incident light into ordinary and extraordinary
components, which then pass through a rotating achromatic half-wave
plate before being recorded simultaneously on the CCD.  Instrumental
polarization of Subaru/FOCAS using the center slit is negligible
($<0.05\%$), as confirmed by our unpolarized standard star exposures.
The 6-hr observations were divided into 9 sets of 10-min exposures at
each of 4 half-wave plate position angles $0^\circ$, $45^\circ$,
$22{\fdg}5$, $67{\fdg}5$.

We used the unpolarized standard star G191B2B to flux calibrate and
remove telluric absorption lines, and used the polarized standard star
HD251204 \citep{tur90} to determine the absolute position angle.  We
reduced individual frames using the FOCAS iraf cookbook, and then
calculated the normalized flux difference for each individual frame:
$F_\theta = (f^o_\theta-f^e_\theta) / (f^o_\theta+f^e_\theta)$.  We
then combined the normalized flux difference for each angle $\theta$
and calculated the Stokes parameters: $Q = 0.5(F_0-F_{45})$ and $U =
0.5(F_{22.5}-F_{67.5})$.  We additionally computed the Stokes
parameters using the method of \citet{mil88}, which accounts for
calibration differences between the ``ordinary'' and ``extraordinary''
light positions, and found no significant differences in the resultant
Q or U.

The degree of polarization is given by $P \equiv \sqrt{Q^2+U^2}$ and
the polarization angle by $\theta \equiv 0.5 \arctan(U/Q)$.  Errors in
$U$ and $Q$ bias the total polarization $P$ in the positive direction,
and we correct for this bias using the simulation-derived correction
of \citet{pat06}.  Galactic interstellar polarization is $<0.2\%$ in
the direction of both targets \citep{hei00} and is therefore
disregarded.

\section{Results}

\begin{figure}
\scalebox{1.2}
{\plotone{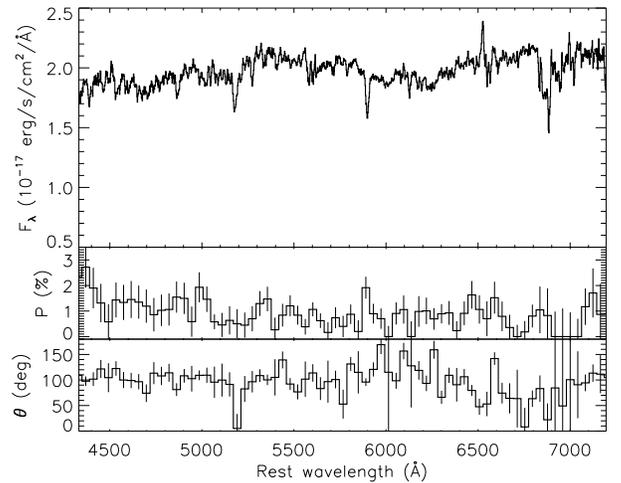}}
\figcaption{Total flux and degree of polarization for 095849+013220,
  the optically dull AGN likely to have a RIAF.  Spectra of
  polarization degree and angle are binned by 20 pixels (41\AA\ in the
  rest frame) to improve their S/N, and the polarization degree is
  corrected for error bias \citep{pat06}.  This target does not show
  any evidence for polarized AGN emission lines but does exhibit
  significant continuum polarization, especially in the blue.  The
  overall mean linear polarization is $0.78 \pm 0.07\%$, and $1.37 \pm
  0.16 \%$ at $4400{\rm \AA} < \lambda < 5050{\rm \AA}$, at an angle
  of $104 \pm 4^\circ$.
\label{fig:riafagn}}
\end{figure}

\begin{figure}
\scalebox{1.2}
{\plotone{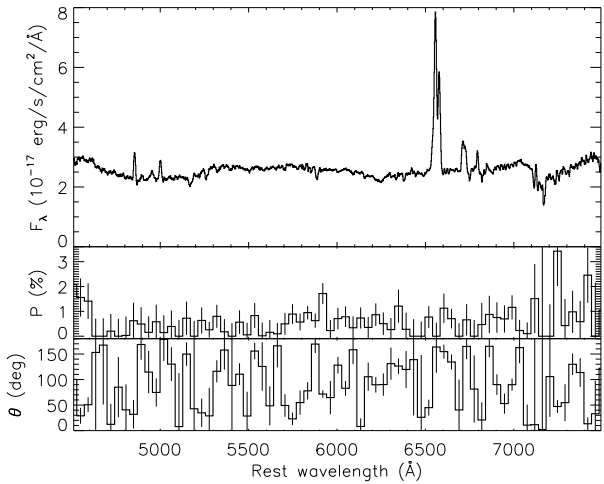}}
\figcaption{Total flux and degree of polarization for 100036+024929,
  the optically dull AGN likely to be a Type 2 AGN diluted by its host
  galaxy.  The Subaru/FOCAS spectrum clearly displays \Ha\ and
  \NII\ and very weak \Hb\ and \OIII.  This is in contrast to our
  previous Magellan/IMACS spectrum, which had much lower S/N and
  showed no emission lines (\Ha\ fell in an IMACS chip gap).  As in
  Figure \ref{fig:riafagn}, the polarization degree is corrected for
  error bias \citep{pat06} and with the polarization angle is binned
  by 20 pixels (43\AA\ in the rest frame).  This AGN does not exhibit
  significant linear polarization, either in emission lines or in the
  continuum.
\label{fig:dilagn}}
\end{figure}

The total flux, degree of polarization, and polarization angle for the
RIAF candidate and dilution candidate AGNs are shown in Figures
\ref{fig:riafagn} and \ref{fig:dilagn}, respectively.  We bin the
degree of polarization and polarization angle by 20 pixels (56\AA\ in
the observed frame) to improve their signal-to-noise.

It is first evident that several narrow emission lines are revealed in
the unpolarized Subaru/FOCAS spectra, as compared to our previous
Magellan/IMACS spectra with much lower S/N.  The RIAF candidate,
095849+013220, may have weak \Ha\ emission, although it is strangely
blueshifted from the absorption line redshift.  The diluted candidate,
100036+024929, shows strong \Ha\ and \NII\ emission (which was missed
in the previous Magellan/IMACS spectrum because of a chip gap) and
weak \Hb\ and \OIII\ emission.

Neither of the polarized fluxes in Figures \ref{fig:riafagn} and
\ref{fig:dilagn} show evidence for reflected AGN emission lines like
\Ha, \Hb, or \OIII.  The \NaI$\lambda$5892 absorption feature might
have increased percent polarization, although it is significant at
only $\sim2\sigma$ in both spectra.  If the increased percent
polarization of \NaI\ is real, it is probably caused by the lower
starlight dilution at that wavelength, and so does not represent an
actual increase in polarized flux.

The RIAF candidate shows significant continuum polarization, while the
diluted candidate does not.  The mean continuum polarization of
095849+013220 is $0.78 \pm 0.07\%$ at an angle of $104 \pm 4\%$
degrees.  The mean continuum polarization in the blue, measured at
$4400<\lambda<5050$\AA, is even stronger: $1.37 \pm 0.16\%$.
Subtracting the host galaxy component, which contributes $\sim$55\% of
the total emission at $\lambda_{\rm rest} \sim 5000$\AA, suggests that
the mean polarization of the AGN component is $\sim$1.7\% overall and
$\sim$3.0\% in the blue.  The polarized continuum roughly doubles from
$\sim$6500\AA\ to $\sim$4500\AA, consistent with a typical unobscured
quasar continuum of $f_{\lambda} \sim \lambda^{-1.6}$ \citep{van01}.

In addition to the two optically dull AGNs from COSMOS, we observed
the cluster BL Lac candidate MS 1455-X2 from \citet{har09} and discuss
its spectropolarimetry results in the Appendix.

\section{Discussion}

Neither of our targets show evidence for polarized broad or narrow
emission lines, which suggests that the lines are not missing due to
anisotropic absorption as in a standard AGN ``torus'' model
\citep{ant93}.  Instead, our RIAF candidate has a blue polarized
continuum.  Here we place limits on polarized emission line flux and
investigate the two possible causes of the continuum polarization:
synchrotron emission or scattering.

\subsection{Limits on Polarized Emission Lines}

For both objects the mean polarization in the wavelength regions of
\Hb\ or \Ha\ broad lines are below or consistent with the measured
continuum polarization in the surrounding regions.  However a
non-detection does not necessarily mean that there are no polarized
emission lines, since very weak lines might be undetected within our
errors.  We approximate each line as a 2000~km/s wide top hat, and
estimate upper limits on the degree of polarization for \Hb\ and \Ha\
broad lines above the polarized continuum using the $3\sigma$ errors
across the line region.  For the RIAF candidate 095849+013220 the
$3\sigma$ upper limits are $2.0\%$ for \Ha\ and $2.1\%$ for \Hb, and
for the diluted AGN 100036+024929 the $3\sigma$ upper limits are
$1.7\%$ for \Ha\ and $3.3\%$ for \Hb.  These upper limits decrease for
broad lines wider than 2000~km/s.  In contrast, ``hidden-BLR'' AGNs
typically have reflected broad emission lines detected at $P \simeq
5\%$ \citep{ant93,bar99,mor00}.

The intrinsic fluxes from both broad lines and continuum are
presumably represented in polarized emission.  For the RIAF candidate
095849+013220 we detect a polarized continuum, and can determine upper
limits in equivalent width (EW) for broad lines in this intrinsic AGN
emission.  The $3\sigma$ upper limits on EW in the intrinsic emission
are ${\rm EW}_{\Ha}<80$\AA\ and ${\rm EW}_{\Hb}<47$\AA.  These limits
are less than those of typical quasars, which have ${\rm EW}_{\Hb}
\sim 50$\AA\ and ${\rm EW}_{\Ha} \sim 200$\AA\ \citep{van01}.

An anisotropic scatterer (e.g., the canonical AGN ``torus'') is not
the reason our two sources lack broad emission lines.  Instead both
AGNs have weaker broad lines than typical Type 1 or hidden-BLR AGNs,
consistent with a RIAF or host galaxy starlight dilution.

\subsection{Continuum Polarization from Synchrotron Emission}

Synchrotron emission from the anisotropic magnetic field associated
with a radio jet has a high degree of linear polarization.  The RIAF
candidate 095849+013220 could be analogous to BL Lacertae AGNs, which
similarly lack both direct and reflected emission lines.  BL Lacs
typically exhibit linear polarization at $\sim2-11\%$, with a median
of $\sim$7\% \citep{jan94,cap07}.  At first glance the $\sim$1.7\%
polarization for the AGN-only component of 095849+013220 is in the low
range of measured polarization for BL Lacs.  However the starlight
dilution in 095849+013220 is not well constrained, and the
polarization of the AGN component might be significantly higher.
Synchrotron emission should extend from radio wavelengths to the
optical/UV as a power-law, and so the resultant polarized flux should
follow the same power-law \citep[e.g.,][]{coh97}.  The observed
increase of polarized flux towards blue wavelengths for 095849+013220
seems to contradict this, but could again be explained by dilution
from an old, red stellar population.  The synchrotron polarization of
BL Lacs is also strongly variable in amplitude and angle
\citep{jan94,coh97}, but we do not have multi-epoch polarization
measurements and so cannot perform this test.

Many of the multiwavelength properties of our RIAF candidate are also
consistent with those of BL Lacs.  The AGN 095849+013220 is
radio-luminous (though it sits just below the canonical radio-loudness
definition): detected at $F_{1.4 \rm GHz}=0.646$~mJy in the VLA-COSMOS
survey \citep{schi10}, it has $f_{1.4~Ghz}/f_{i,AGN}=6.7$.  This ratio
is a factor of a few lower than typical radio to optical ratios for BL
Lacs \citep[e.g.,][]{sto90}, but again this can be explained if our
starlight dilution estimate is too low.  X-ray selected BL Lacs (HBLs)
have similarly high $X/O$ ratios to 095849+013220, although the
typical $X/O$ of all BL Lacs is lower and more like that of the
typical quasar population \citep{plo08}.  The polarized continuum adds
another piece of evidence that 095849+013220 \citep[and possibly other
optically dull AGNs with similarly high radio/optical and
X-ray/optical ratios from][]{tru09c} might be unified with BL Lacs as
another kind of misaligned FR I radio galaxy \citep{cap07}.

\subsection{Continuum Polarization from Scattering}

Scattering also causes continuum polarization, as the observer sees
light reflected off the scattering surface.  The polarized continuum
of 095849+013220, while blue, actually has the same $f_{\lambda} \sim
\nu^{-1.6}$ slope as an unobscured quasar continuum.  This slope
suggests that optically thick clumpy dust clouds are responsible for
the scattering.  In contrast the wavelength dependence of optically
thin dust scattering would cause a much bluer slope
\citep[e.g.][]{tran95}, and electron scattering would require a much
larger gas mass than indicated by the X-ray column density \citep[$N_H
  \simeq 4 \times 10^{20}$~cm$^{-2}$,][]{tru09c}.  Scattering by
clumpy dust is also thought to cause the polarized continuum observed
in many Type 1 and Type 2 AGNs \citep{ant85,ogl99,kis01}.

The canonical AGN ``torus'' is thought to be made of clumpy dust
\citep{nen08}, and so could be the scatterer.  The AGN unified model
invokes this torus to block and reflect the broad emission lines
\citep{ant93}.  In 095849+013220, however, we observe a polarized
continuum but no polarized emission lines.  Since the broad emission
line region is $<<1$~pc from the continuum \citep{elv00,pet06}, it is
not physically plausible for any scattering surface to reflect the
continuum but not the emission lines.  For this reason we interpret
the detection of a polarized continuum without broad lines as evidence
for a very weak or nonexistent BLR, as expected for radiatively
inefficient accretion \citep{tru11}.

\section{Conclusions}

We present Subaru/FOCAS spectropolarimetry for two optically dull
AGNs: the RIAF candidate 095849+013220 and the galaxy-diluted Type 2
AGN candidate 100036+024929.  Neither source exhibits reflected broad
emission lines, with 3$\sigma$ upper limits of $\lesssim$2\% polarized
broad lines.  This rules out simple torus obscuration as the reason
behind the lack of broad emission lines.  Despite the lack of
polarized broad line emission, in the RIAF candidate we additionally
observe a polarized continuum.  The blue spectral slope of the
polarized flux suggests that it represents the intrinsic AGN
continuum, either observed directly as synchrotron emission or as
reflected by clumpy dust.  Because we observe the direct emission from
the central AGN engine without broad emission lines, we suggest that
this source has an intrinsically weaker or nonexistent BLR.  The
spectropolarimetry is another piece of evidence that the optically
dull AGN 095849+013220 has a radiatively inefficient accretion flow.

\acknowledgements

We thank Takashi Hattori and the staff of Subaru for excellent support
during and after observations.  We thank Kenta Matsuoka for help in
reducing the FOCAS data.  Masaomi Tanaka was invaluable in
interpreting the spectropolarimetry results.  We also thank Hien Tran
and Makoto Kishimoto for help in interpreting a polarized continuum in
AGNs.  JRT acknowledges support from NSF/DDEP grant \#0943995 and NSF
grant \#AST-0808133, and with CDI acknowledges support from
\#AST-0908044.  FC acknowledges support from the Blancheflor
Boncompagni Ludovisi foundation and the Smithsonian Scholarly Studies.
BCK acknowledges support from NASA through Hubble Fellowship grant
\#HF-51243.01 awarded by the Space Telescope Science Institute, which
is operated by the Association of Universities for Research in
Astronomy, Inc., for NASA, under contract NAS 5-26555.  YT
acknowledges support from JSPS grants \#17253001 and \#19340046.

\appendix
\section{Observations of the cluster BL Lac candidate MS 1455-X2}

\citet{har09} reported that the cluster object MS 1455-X2 is luminous
in X-ray and radio emission and yet lacks emission lines in its
optical spectrum.  Its basic properties are given in Table
\ref{tbl:odagns}.  Its optical colors indicate emission blueward of
4000\AA\ in excess of that expected for a lineless red galaxy, and so
MS 1455-X2 is similar to the RIAF candidate we observe in COSMOS.
Indeed, \citet{har09} suggest that it is a low-luminosity BL Lac, and
predict that such objects may be common in clusters and can provide
the feedback necessary to heat cluster cores.

We observed MS 1455-X2 for a total of 160 minutes, divided into 4 sets
of 10-min exposures at each of the 4 half-wave plate position angles.
We used the same instrumental parameters and reduction process as
described in Section 2.  The shorter exposure time and faintness of
the target ($r=20.05$) mean that we cannot put meaningful limits on
the amount of emission line polarization for MS 1455-X2.  However we
detected a significant total polarization of $P=2.57 \pm 0.47\%$ at
$4800<\lambda_{\rm rest}<7700$\AA.  This high degree of polarization
strongly supports the hypothesis that this source is a low-luminosity
BL Lac AGN, especially considering some amount of starlight dilution
from the host galaxy.

\end{document}